\documentclass[article]{aa}

\usepackage[varg]{txfonts}
\usepackage{amsmath}
\usepackage{graphicx}
\usepackage{hyperref}
\usepackage{natbib}

\bibpunct{(}{)}{;}{a}{}{,}

\begin{document}

\title{Core and crust contributions in overshooting glitches: the Vela pulsar 2016 glitch}

\author{P.~M.~Pizzochero\inst{\ref{inst1}, \ref{inst2}}\thanks{pierre.pizzochero@mi.infn.it}
	\and 
	A.~Montoli\inst{\ref{inst1}, \ref{inst2}}\thanks{alessandro.montoli@unimi.it}
	\and
	M.~Antonelli\inst{\ref{inst3}}\thanks{mantonelli@camk.edu.pl}
}

\institute{
	Dipartimento di Fisica, Universit\`a degli Studi di Milano, Via Celoria 16, 20133 Milano, Italy\label{inst1}
	\and
	Istituto Nazionale di Fisica Nucleare, sezione di Milano, Via Celoria 16, 20133 Milano, Italy\label{inst2} 
	\and
	Nicolaus Copernicus Astronomical Center of the Polish Academy of Sciences, Bartycka 18, 00-716 Warszawa, Poland\label{inst3}
}

\abstract{
During the spin-up phase of a large pulsar glitch - a sudden decrease of the rotational period of a
neutron star - the angular velocity of the star may overshoot, namely reach values greater than that
observed for the new post-glitch equilibrium. These transient phenomena are expected on the basis
of theoretical models for pulsar internal dynamics and their observation has the potential to provide
an important diagnostic for glitch modelling. In this article we present a simple criterion to assess
the presence of an overshoot, based on the minimal analytical model that is able to reproduce an
overshooting spin-up. We employ it to fit the data of the 2016 glitch of the Vela pulsar, 
obtaining estimates of the fractional moments of inertia of the internal superfluid components involved in the glitch, of the rise and decay  timescales of the overshoot, and of the mutual friction parameters between the superfluid components and the normal one. We study the cases with and without strong entrainment in the crust: in the former, we find indication of a large inner core strongly coupled to the observable component and of a reservoir of angular momentum extending into the core to densities below nuclear saturation, while in the latter a large  reservoir extending above nuclear saturation and a standard normal component without inner core are suggested.
}


\maketitle

\section{Introduction}
\label{sec:intro}

Radio pulsars are known for their stable rotational period. Nevertheless, several pulsars exhibit sudden and sporadic spin-up events of small amplitude, known as glitches \citep{espinoza+2011}.
Since the pioneering work of \citet{baym+1969}, several models proposed to describe glitches by formally dividing the spinning neutron star into two parts: a normal component, corotating with the observed beamed radiation of magnetospheric origin, and a superfluid neutron component \citep{haskellmelatos2015}. 
A difference of angular velocity may develop between the two components (constituting a reservoir of angular momentum) thanks to the pinning of the superfluid vortices to impurities of the crustal lattice \citep{andersonitoh1975} or to the quantised flux-tubes of magnetic field permeating  the outer core \citep{guerci_alpar_2015}.
Following this paradigm, several models have been employed to study glitching pulsars, yielding indirect constraints on the neutron star structural properties through observations \citep{dattaalpar1993,LE99, andersson+2012, chamel2013,newton_etal15,ho+2015,pizzochero+2017,montoli+2020}.

The possibility to test our understanding of the glitch mechanism is hindered by the difficulty to observe  glitches in the act. In fact, glitch rises are generally not resolved, due to intrinsic noise in the time of arrival of single pulsations. 
Moreover, in spite of the fact that the Vela pulsar has been monitored for fifty years, only a couple of notable events allowed to put an upper limit of $40$ s on the timescale of the glitch spin-up \citep{dodson+2002, dodson+2007}.
Only recently, with the observation of a glitch on 12th December 2016, it has been possible to measure the time of arrival of single pulses during the glitch with a precision never achieved before, and thus to obtain some information on the first seconds after the event \citep{palfreyman+2018}.   In particular, a new strong upper limit of $12.6$ s on the timescale of the glitch spin-up has been determined by \citet{ashton+2019}.
This kind of observation opens a new window for theoretical speculations. 
In fact, complex behaviour  during the spin-up and the first minute of the post-glitch relaxation has been predicted in simulations based on hydrodynamical models of the neutron star internal structure, when more than just two rigid components are considered \citep{haskell+2012, anto_haskell_2013, antonellipizzochero2017, graber+2018}:
when the superfluid component is allowed to sustain non-uniform rotation, different regions may experience different friction and hence recouple to the observable normal component on different timescales, giving different glitch shapes.

In particular, depending on the strength of the couplings and on the initial conditions for the relative motion between the various components, a glitch overshoot (a transient interval in which the  observable component spins at a higher rate  than  the post-glitch equilibrium value,  obtained by emptying the whole angular momentum  stored into the superfluid reservoir \citep{antonellipizzochero2017} is observed in such models.

Two recent studies have already used the data from the 2016 glitch: in  \citet{graber+2018}, the drag between the charged crust and the crustal and core superfluids has been constrained; in \citet{ashton+2019}, different phenomenological models have been compared to the timing results, obtaining the best current  limits on the glitch rise timescale. Both studies also confirmed the presence of an overshoot.

In this article we first give a simple quantitative result for the onset of a glitch overshoot, by employing a three-rigid-component model for the glitch dynamics (which is the \emph{minimal} model capable of reproducing an overshoot). We then use the model to fit the 2016 glitch of the Vela pulsar.
The advantage of the present treatment is that it provides an analytical form for the timing residuals, which is directly related to physical parameters of the neutron star. Indeed, in addition to determining  the rise and decay timescales of the overshoot,
the fit results allow to derive some  physical properties of the three components, like  the moment of inertia fractions and  the drag parameters between the two superfluid components and the normal one.

\section{Three component model}
\label{sec:model}

Generalising the approach of  \citet{baym+1969}, the pulsar is described by means of three rigidly rotating components. 
We consider two neutron superfluid components (labelled with 1, 2 subscripts), that exchange angular momentum with a normal component $p$ on timescales $\tau_{1, 2}$. The $p$-component is interpreted as all the charged particles coupled to the observable magnetosphere on timescales shorter than $\tau_1$ and $\tau_2$, while we do not need to specify what the two superfluid components are: in fact, the equations we are going to write are completely general, as they derive from conservation of angular momentum and the only assumption of rigid rotation of the three components. Physically, however, they could represent  the P-wave superfluid in the core and the S-wave one the crust, as the physical conditions of these regions  are completely different. We thus employ a set of three equations, one for the conservation of angular momentum, and two representing the interaction between the normal component and each of the two superfluid components:
\begin{align}
\begin{split}
\dot{\Omega}_p &= - \frac{1}{x_p} \left( x_1 \dot{\Omega}_1 + x_2 \dot{\Omega}_2 + |\dot{\Omega}_\infty| \right)
\\
\dot{\Omega}_1 &= -x_p \frac{\Omega_1 - \Omega_p}{\tau_{1}} =  -x_p \frac{\Omega_{1p}}{\tau_{1}} 
\\
\dot{\Omega}_2 &= -x_p \frac{\Omega_2 - \Omega_p}{\tau_{2}} =  -x_p \frac{\Omega_{1p}}{\tau_{1}} 
\end{split}
\label{eq:3c}
\end{align}
where $\Omega_{ip}= \Omega_i - \Omega_p$ ($i=1, \, 2$) are the lags and where $x_i=I_i/I$ ($i=1, \, 2, \, p$) are the ratios of the partial moment of inertia $I_i$ of the $i$-component with respect to the total one $I=I_1+I_2+I_p$, so that $x_1+x_2+x_p=1$.
The quantity $|\dot{\Omega}_\infty|$ sets the intensity of the external braking torque (for Vela $\dot{\Omega}_\infty \approx -9.78\times 10^{-11}$ rad/s${^2}$, but its precise value is unimportant in the following analysis).
Equation \eqref{eq:3c} is valid without superfluid entrainment: this will be discussed in a dedicated section. 

We remark that a three component model was also introduced in \citet{graber+2018}, but with a  differential rotation associated to  the reservoir: this was necessary to study the density-dependent drag parameters, but it requires a numerical integration of the dynamical equations. The model in equation \eqref{eq:3c}, to which the equations in  \citet{graber+2018} reduce when imposing rigid rotation and constant drag, is the simplest analytical treatment which can reproduce an overshoot, allowing to derive directly the average properties of the superfluid components (fractional moment of inertia and average drag).

For a real pulsar we expect the two timescales $\tau_{1,2}$ to be complicated functions of the instantaneous angular velocity lags $\Omega_{ip}= \Omega_i - \Omega_p$ and also to depend on the past history of the vortex configuration and internal stresses. In a model with rigid components, these timescales define the strength of the vortex-mediated mutual friction, which is responsible for the angular momentum exchange, suitably averaged over the region of interest.
To better compare with \citet{graber+2018}, the timescales $\tau_{i}$ can be connected to the large-scale hydrodynamic mutual friction coefficients $\mathcal{B}_{i}$. In turn, these are related to the dimensionless drag parameters $\mathcal{R}_{i}$ (which are the results of theoretical calculations) by the relation $\mathcal{B}_i=\mathcal{R}_i/(1+\mathcal{R}_i^2)$, see e.g. \cite{anderssonMF2006}. 
The dynamical equations for  rigidly rotating superfluids in the presence of mutual friction are \citep[see e.g. ][]{haskellmelatos2015}: 
\begin{align}
\dot{\Omega}_i=-2 \Omega_i \mathcal{B}_i (\Omega_i - \Omega_p) \simeq -2 \Omega_p(0) \mathcal{B}_i (\Omega_i - \Omega_p) \ \ \ \  \ \ (i=1,2).
\label{eq:drag equation}
\end{align}
where we approximated $\Omega_1 = \Omega_2 = \Omega_p(0)$ in the prefactor, since the lags between the superfluids and the normal component are always  orders of magnitude smaller than the  angular velocity of the normal component. Comparing equations \eqref{eq:3c} and \eqref{eq:drag equation} we can write the (approximate) relation:
\begin{align}
\mathcal{B}_i =  \frac{x_p}{2 \Omega_p(0) \tau_i}  \ \ \ \  \ \ (i=1,2).
\label{eq:timescales}
\end{align}
In the following, we will take a nominal value $\Omega_p(0)=70.29$ rad/s \citep{dodson+2002}.

How to construct realistic models of vortex-mediated mutual friction (i.e. understanding the many-vortex dynamics in neutron stars in the presence of pinning sites) is one of the current challenges of glitch theory.
In the present phenomenological description we assume  that, at the glitch time, $\tau_{1,2}$ undergo a transition from  large ``pre-trigger'' values to much smaller ``post-trigger'' values: the nature of the trigger is undetermined but, according to this simple picture, the vortices change their state of motion, increasing their creep rate and thus mimicking the  onset of a vortex avalanche \citep{Alpar84a}. 
In fact, if the vortices of the $i$-component are pinned, or their motion is severely hindered, the timescale $\tau_i$ diverges, so that the corresponding $\Omega_i$ remains constant regardless of the state of motion of the other components. 
Therefore, pinning implies the decoupling of that component from the rest of the system, while a suddenly  recoupling of such a component results in an exchange of angular momentum from the superfluid to the crust,  leading to a glitch. 

We now study the solutions of the system \eqref{eq:3c}. Since the main goal of the present analysis is to provide the simplest criteria for overshooting glitches, we take $\tau_1$ and $\tau_2$ to be constants for $t>0$, thus neglecting the repinning process (that may be nonetheless important for a complete description of the inter-glitch dynamics): this approximation should hold at least for the overshoot phase.
 
First, it is useful to rewrite the problem using the lags $\Omega_{ip} $ as variables instead of $\Omega_{1}$ and $\Omega_{2}$. 
In this way, the two equations for $\dot{\Omega}_{ip}$ do not depend on $\Omega_p$, and it is possible to solve them independently from the equation for $\dot{\Omega}_p$. 
To set the unknown initial conditions $\Omega_{ip}^0$ we rely on a physical assumption: we impose the component 1 to be a ``passive'' one that does not change its creep rate (i.e. $\tau_1$ is always constant and $\Omega_{1p}^0= \tau_1 \,|\dot{\Omega}_\infty|/x_p$), while the component 2 (acting as the reservoir) has a lag $\Omega_{2p}^0=\omega(0)+\tau_2 \,|\dot{\Omega}_\infty|/x_p$. The positive quantity $\omega(0)$ is the excess lag with respect to the asymptotic post-glitch steady-state lag, which has been accumulated before the triggering event.

The angular velocity of the normal component for $t>0$ can finally be written as $\Omega_p(t) = \Omega_p^0+\dot{\Omega}_\infty t + \Delta\Omega_p(t)$, where the difference  with respect to the steady-state, $\Delta\Omega_p(t)= \Omega_p(t) -\Omega_p^0-\dot{\Omega}_\infty t$, is given by
\begin{equation}
\Delta \Omega_p (t) 
= 
\Delta \Omega_p^\infty - \epsilon \frac{Q + R}{\tau_+} e^{-t / \tau_-} + \epsilon \frac{Q -R}{\tau_-} e^{-t/\tau_+} \, ,
\label{eq:solution}
\end{equation}
where $\Delta \Omega_p^\infty  = x_2 \omega(0)$. In the above expression the following constants have been defined:
\begin{align}
Q &= 1 - x_1 - \beta (1 - 2x_1 - x_2)\\
R &= \sqrt{[1 - x_1 + \beta (1 - x_2)]^2 - 4 \beta(1 - x_1 - x_2) } \\
\epsilon &= \frac{x_2 \tau_1}{2(1-x_1 - x_2) R}\, \omega(0)\\
\tau_\pm &= \frac{2 \beta \tau_1}{1 - x_1 + \beta(1 - x_2) \mp R}\, ,
\label{eq:definitions}
\end{align}
where $\beta = \tau_2/\tau_1$ is the ratio between the two timescales and  $\tau_+ > \tau_- >0$.

We remark that our solution, equation \eqref{eq:solution}, has the same form of one of the phenomenological models studied in \citet{ashton+2019}. Here we make the further step of connecting the coefficients of the two exponentials and the timescales  $\tau_-$ and $\tau_+$ to physical parameters of the neutron star, that can thus be inferred after comparison with the data.

It is interesting to point out some properties of the expression in \eqref{eq:solution}. 
First, we have $\Delta \Omega_p \rightarrow \Delta \Omega_p^\infty $ for $t \rightarrow \infty$, which is the glitch amplitude measured at $t\gg \tau_+$. 
The glitch amplitude, however,  can overcome this asymptotic value at earlier times: the presence of such an overshoot is revealed by the existence of a maximum in $\Delta \Omega_p(t)$, occurring at time
\begin{equation}
t_{\rm max} = \frac{\tau_1 \beta}{R} \ln\left(\frac{Q + R}{Q - R} \right).
\label{eq:tmax}
\end{equation} 
The maximum exists only if the argument of the logarithm is positive, which implies $\beta < 1$. 
In other words, the condition for an overshoot is that the post-glitch timescale $\tau_2$ associated to the ``active'' component (that in the pre-glitch state was only loosely coupled to the rest of the star) must be smaller than the timescale $\tau_1$ of the ``passive'' component (that does not change its coupling). 
From the physical point of view, the overshoot occurs if the ``active'' superfluid region that stores the angular momentum for the glitch can transfer   its  excess of angular momentum to the normal component faster than the typical timescale the ``passive'' superfluid component reacts with.
This behaviour can already be seen in Figure 3 of \citet{graber+2018} and was explicitly noted in \citet{ashton+2019}. Here we confirm this, by giving it a mathematical foundation.

\begin{figure}
	\centering
	\includegraphics[width=.45\textwidth]{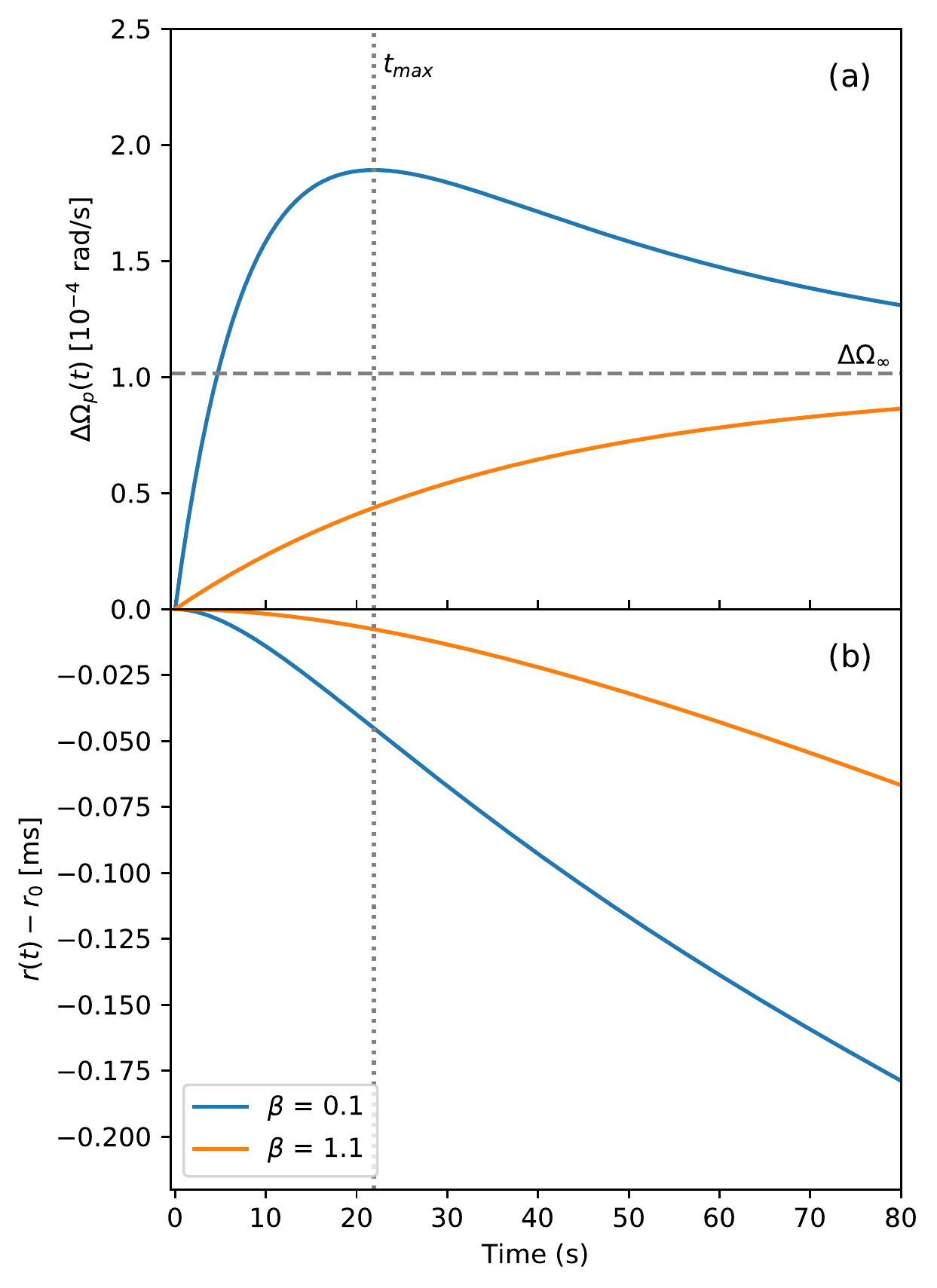}
	\caption{Representation of a glitch with ($\beta = 0.1$) and without ($\beta = 1.1$) overshoot. The remaining parameters are taken from Table~\ref{tab:results}. In the upper panel we show the angular velocity with respect to the steady state, $\Delta \Omega_p(t)$,  in the lower panel the (shifted) residuals, $r(t)-r_0$. The flex  in the residuals for the glitch with overshoot is marked by the intersection with the vertical line at $t=t_{\rm max} $.}
	\label{fig:overshoot}
\end{figure}

Following \cite{graber+2018}, we now study the time dependence of the time residuals $r(t)$ with respect to the timing model of a uniformly decelerating pulsar:
\begin{equation}
r(t) \, = \, r_0 \, - \, \frac{1}{\Omega_p(0)} \int_0^t \Delta \Omega_p(t') \, \mathrm{d}t' \,\, ,
\label{eq:res}
\end{equation}
where a constant residual $r_0$ has been added to account for an offset due to magnetospheric changes.
It is easy to see how the condition for the overshoot is translated in terms of the residuals: the glitch presents an overshoot if $r(t)$ is first concave downwards, then upwards, with a flex point at $t = t_{\rm max}$. 
Conversely, a non-overshooting glitch is always concave downwards. 
 In Figure~\ref{fig:overshoot} we show the behaviour of both the angular velocity with respect to the steady state and of the residuals, for  $\beta=0.1<1$ (overshoot) and  for $\beta=1.1>1$ (no overshoot).

Looking at the averaged data for the 2016 Vela glitch shown in Figure 5 of \citet{graber+2018}, we deduce that that glitch presents an overshoot, as it shows a positive concavity before reaching steady-state. 
The first instants of negative concavity are lost, probably due to the extremely fast acceleration of the star and to the magnetospheric change in the pulsar magnetic field \citep{palfreyman+2018},
although a flex can be detected (with difficulty, due to the scale of the figure) in the solid line  a few seconds after the beginning of the glitch. The overshoot was also recently confirmed by \citet{ashton+2019}.

\begin{table}
\caption{Results of the fit for the 7 independent parameters of  equation~\eqref{eq:res}.  The time of beginning of the glitch,  $t_{gl}$, is given with respect to $t_0$, while $t_{\rm max} $ is given with respect to $t_{gl}$. The relative error on $\Delta \Omega_p^\infty$ is of order $10^{-5}$, while the other errors are at $1\sigma$ confidence level. The second part of the table reports some dependent quantities and their propagated errors, obtained from  equations~\eqref{eq:timescales}, \eqref{eq:definitions} and \eqref{eq:tmax}.}
\label{tab:results}
\centering
\begin{tabular}{l @{\hspace{0.1\textwidth}} r}
\hline \hline
Parameter & Value\\
\hline
$x_1$ & $0.60 \pm 0.01$\\
$x_2$ & $0.08 \pm 0.01$\\
$\tau_1$ & $34.6 \pm 1.3$ s\\
$\beta$ & $(2.3 \pm 1.6) \cdot 10^{-3}$\\
$\Delta \Omega_p^\infty$ & $1.014 \cdot 10^{-4}$ rad/s\\
$r_0$ & $0.086 \pm 0.002$ ms\\
$t_{gl}$ & $ 2.0 \pm 0.1$ s\\
\hline
$x_p$ & $0.32 \pm 0.02$  \\
$\tau_2$ & $0.08 \pm 0.06$ s\\
$\tau_-$ & $0.20 \pm 0.14$ s\\
$\tau_+$ & $43.3 \pm 2.1$ s\\
 $t_{\rm max}$& $1.2 \pm 0.7$ s\\
$\mathcal{B}_1$ & $(6.6 \pm 0.6) \cdot 10^{-5}$ \\
$\mathcal{B}_2$ & $(2.8 \pm 2.2) \cdot 10^{-2}$ \\
\hline
\end{tabular}
\end{table}

\section{Fit to the 2016 Vela glitch}
\label{sec:fit}

We now fit expression \eqref{eq:res}  (which contains 7 independent parameters) to the data of the residuals made available by \citet{palfreyman+2018} using a least-squares method.
However, some precautions have to be taken. First, although the glitch time $t_{gl}$ and amplitude $\Delta \Omega_p^\infty $ were already estimated by \citet{palfreyman+2018}, here we will take them as free parameters, thus allowing for a check of our results.
Secondly, as noticed by \citet{palfreyman+2018}, soon after a null (missing) pulse  at  time $t_0$, a sudden and persistent increase of the timing residuals has been detected in the time interval between $t_1=t_0+1.8$s and  $t_2=t_0+4.4$s (cf. Figure~\ref{fig:data} for the relative positions of these times).
This behaviour  may correspond to a slow down of the star just before the glitch \citep{ashton+2019} or to a magnetospheric change in the star \citep{palfreyman+2018}. As we are not able to model this kind of phenomena with the current equations, we will just consider the resulting positive offset in the timing residuals $r_0$ as a variable for our fit. For the same reason, we will have to neglect some of the data after the occurrence of the glitch. Indeed, during the interval $\Delta t _m=t_2 - t_1$ the emitting magnetosphere has decoupled from (is not corotating with) the rapidly accelerating  crust:   the persistent positive offset in the mean of the timing residuals and their associated low variance observed by \cite{palfreyman+2018} during $\Delta t_m$ cannot describe the overshooting normal component, which instead would correspond to decreasing residuals. Therefore, the data around the interval $\Delta t _m$ do not describe the crust rotation and should be excluded from the fit.

\begin{figure}
	\centering
	\includegraphics[width=.45\textwidth]{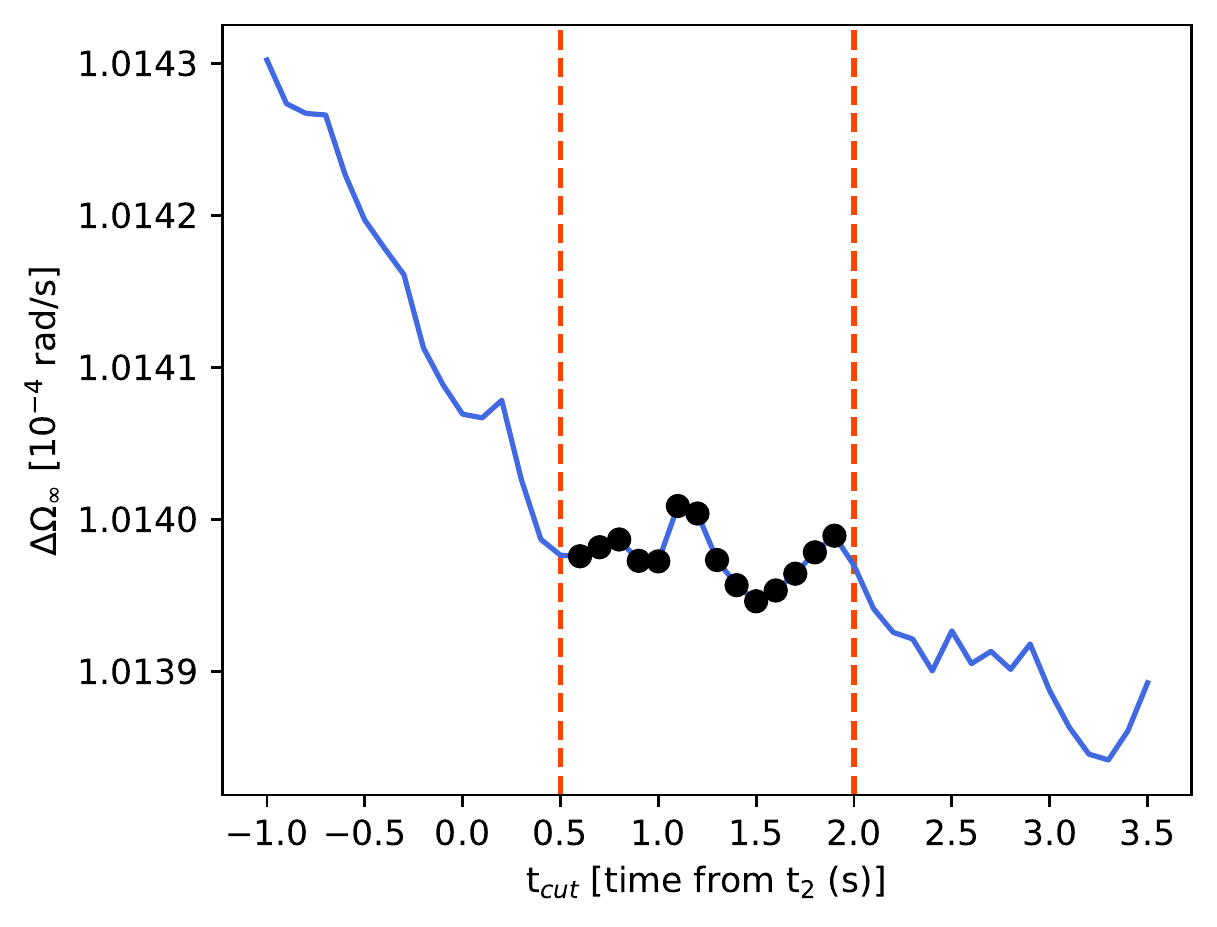}
	\caption{We show the results of the fit for the parameter $\Delta \Omega_p^\infty $ as a function of $t_{\rm cut}$, the time (measured with respect to $t_2$) before which we neglect the data. We vary $t_{\rm cut}$ by steps of 0.1s, and connect the results by a line for clarity. The vertical lines define the region we have chosen to evaluate $\Delta \Omega_p^\infty $; the mean and standard deviation reported in Table~\ref{tab:results} are taken for the values of $\Delta \Omega_p^\infty $ marked by black dots.}
	\label{fig:dOmega}
\end{figure}

\begin{figure}
	\centering
	\includegraphics[width=.45\textwidth]{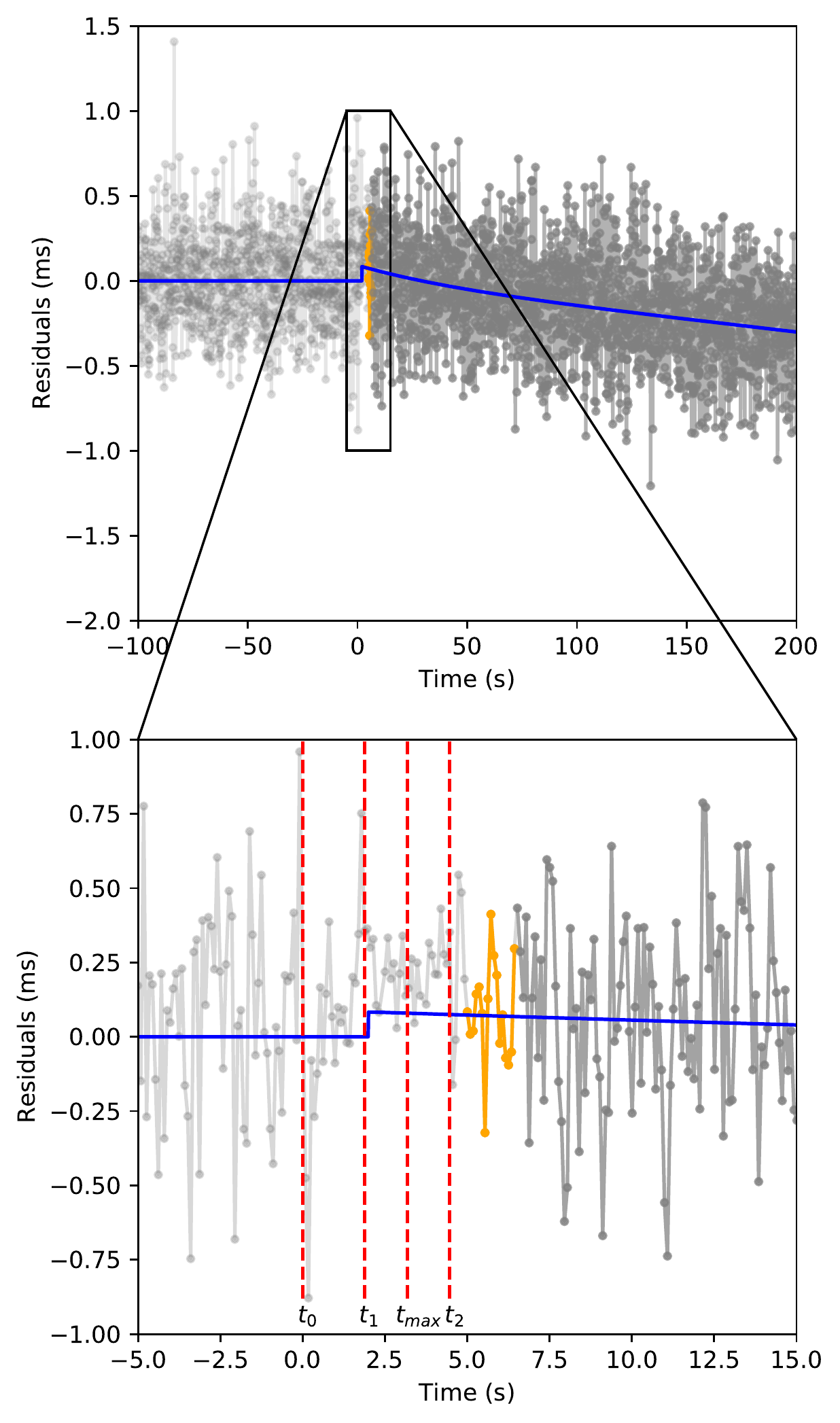}
	\caption{The timing residuals around the time of the glitch, as obtained in \citet{palfreyman+2018}. Superimposed in blue, we plot our best fit for the residuals (equation \eqref{eq:res} with the parameters of  Table~\ref{tab:results}). In the zoom we indicate the times $t_0, t_1, t_2$ defined in \citet{palfreyman+2018} and our result for $t_{\rm max} $ (cf. Figure~\ref{fig:glitch}). The glitch begins right after $t_1$. The data points are connected by a line for clarity: in light grey those always omitted from the fit, in dark grey those always included, in orange the region  corresponding  to the interval of $t_{\rm cut}$ over which we evaluate the parameters of the model, as explained in the text (cf. Figure~\ref{fig:dOmega})}
	\label{fig:data}
\end{figure}

 In order to decide  how much data to neglect, we proceed as follows:  
defining $t_{\rm cut}$ as the time before which the data are neglected, we perform the fit varying $t_{\rm cut}$ between $t_2-1$s and $t_2+4$s by steps of 0.1s (the frequency of the Vela being about 11Hz, this amounts to eliminating one  data point at each successive fit). The fitted parameters can then be plotted as a function of $t_{\rm cut}$: in Figure~\ref{fig:dOmega} this is shown for $\Delta \Omega_p^\infty $ (the best determined parameter in our model, due to the extension of the data well after relaxation has completed). The fitted $\Delta \Omega_p^\infty $ first decreases until $t_{\rm cut}=t_2+0.5$s, then stabilises until $t_{\rm cut}=t_2+2$s, then decreases to stabilise at a slightly smaller value until $t_{\rm cut}=t_2+3$s. Short after that, the fitting  of the data with expression \eqref{eq:res}, containing two exponentials,  does not converge anymore, probably because too much data has been omitted to resolve the short time component and determine its parameters. The variations of $\Delta \Omega_p^\infty $ even during the 'stable' phases shows the sensitivity of our fit to the choice of data range: even removal of one data point affects the result, which reflects the inherent noise in the timing residual data. We then decide to take as final result for each parameter the mean and standard deviation calculated from the  values it assumes when $t_{\rm cut}$ varies in the interval $[t_2+0.5$s, $t_2+2$s]. We have also checked that taking the mean and standard deviation in the longer interval $[t_2+0.5$s, $t_2+3$s] yields mean values within the previous errors and larger standard deviations (as obvious from the figure for $\Delta \Omega_p^\infty $). However, we prefer to adopt the smaller interval (whose datapoint are marked in orange in Figure~\ref{fig:data}), which eliminates less information about the short time component.

Although not compelling, the fact that $\Delta \Omega_p^\infty $ 'stabilises' only five pulsar revolutions after $t_2$ seems to indicate that shortly after $\Delta t_m$ the magnetosphere recouples with the normal component. To our knowledge, no theoretical work on the decoupling and recoupling of  the magnetosphere following a glitch has been performed, so that the timescale of order  $\Delta t_m=2.6$ s for the duration of this process remains, at present, only speculative. Incidentally, the recent work by \citet{quakevela2016} studies the response of the magnetosphere to a quake in the crust, arguing that this is the cause of the null pulse at $t_0$ and speculating that the quake may be the trigger of the glitch.

\begin{figure}
	\centering
	\includegraphics[width=.45\textwidth]{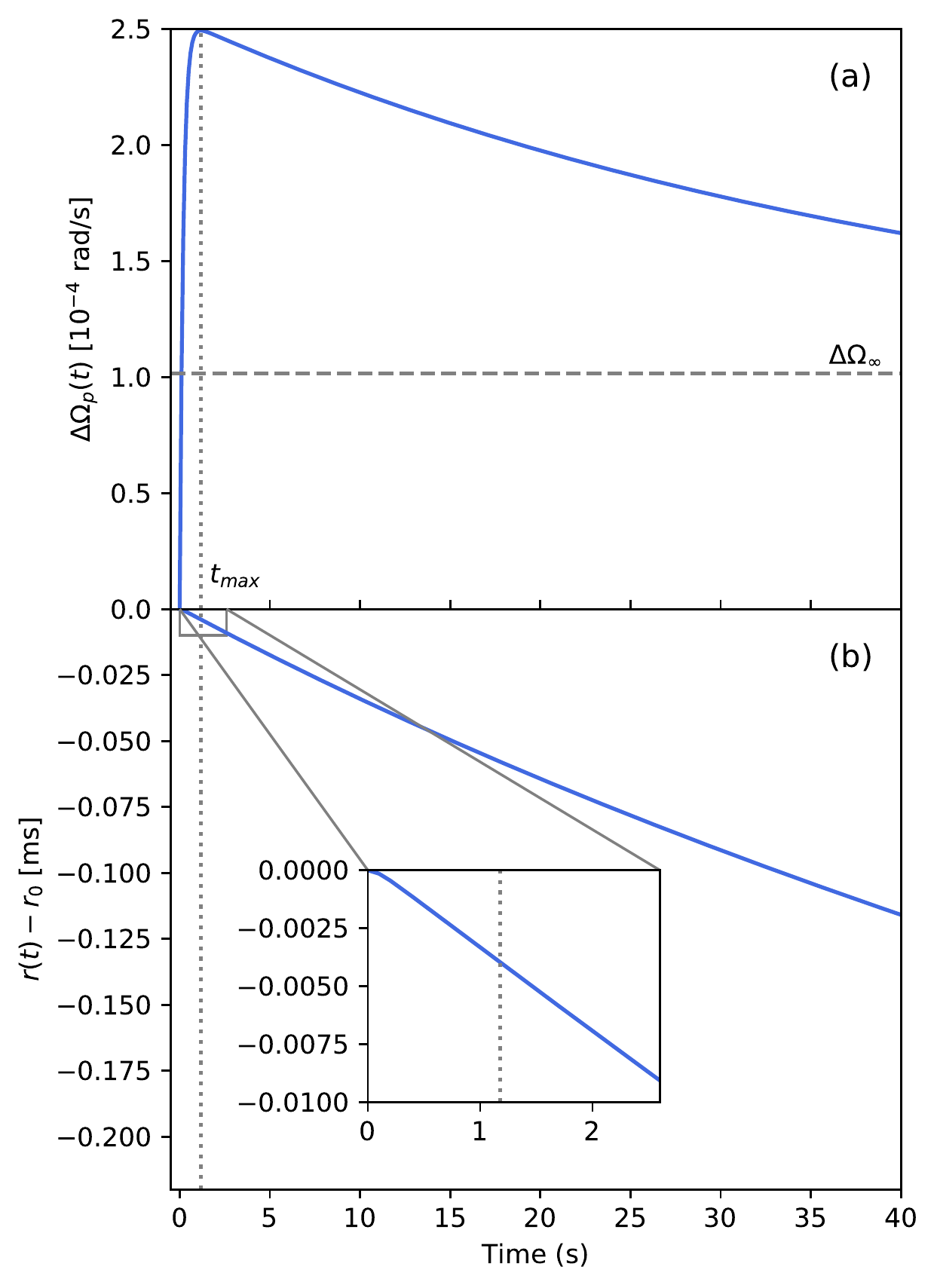}
	\caption{ In the upper panel we show the angular velocity with respect to the steady state, $\Delta \Omega_p(t)$,  in the lower panel the (shifted) residuals, $r(t)-r_0$, corresponding to the values of the fitted parameters in Table~\ref{tab:results}.  The flex  in the residuals 
	 is marked by the intersection with the vertical line at $t_{\rm max} $ (see the zoom) and the time is measured from the beginning of the glitch.}
	\label{fig:glitch}
\end{figure}

\begin{figure}
	\centering
	\includegraphics[width=.45\textwidth]{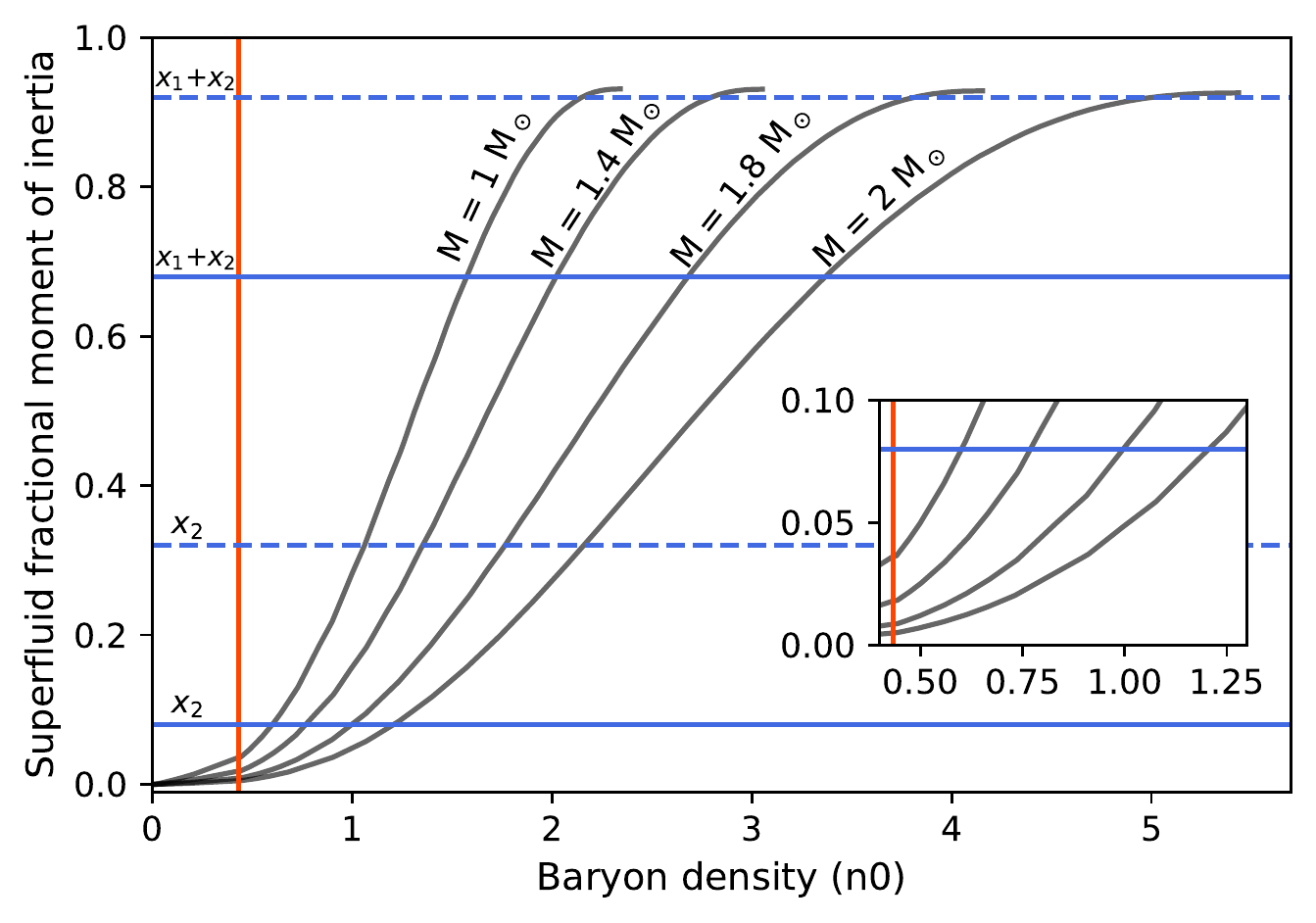}
	\caption{
	Moment of inertia fraction of the neutrons enclosed in a spherical shell extending from the radius at which the neutron drip starts to the radius corresponding to a certain baryon number density. The baryon density corresponding to the internal boundary of the shell is given in units of $n_0$ (nuclear saturation). The vertical line marks the core-crust transition at $0.45 n_0$. The horizontal line represent $x_2$ and $x_1+x_2=1-x_p$ without (solid lines: $m^*_1=m^*_2=1$) and with strong entrainment in the crust (dashed lines: $m^*_1=1, \ m^*_2=4$). We used the SLy4 equation of state and four reference masses: 1, 1.4, 1.8 and 2 $M_\odot$. The inset is a zoom on the outermost regions of the core.	}
	\label{fig:MoI}
\end{figure}

The data points were taken from \citet{palfreyman+2018} and they cover  72 min across the glitch: part of them is shown in Figure~\ref{fig:data}.  The results for the 7 independent parameters of  equation~\eqref{eq:res} are reported in Table~\ref{tab:results}; in its lower part, we also show some dependent quantities, that can be derived from the equations in the previous section. The glitch, $\Delta\Omega_p(t)$, and its (shifted) residuals, $r(t)-r_0$,   corresponding to the parameters in the table are  shown in Figure~\ref{fig:glitch}, while the curve for the residuals $r(t)$ is  also superimposed to the data in Figure~\ref{fig:data}.
 
 The results of  Table~\ref{tab:results} yield some interesting considerations. First of all,  the glitch size $\Delta \Omega_p^\infty$ is the same as what obtained in \citet{palfreyman+2018} ($\Delta \Omega_p^\infty= 1.006 \cdot 10^{-4}$ rad/s) once their long-term ($\tau_d=0.96$ day) decay term  $\Delta \Omega_d=0.008\cdot 10^{-4}$ rad/s (absent in our model, since the data we use extend to about 34 minutes after the glitch time) has been added.

Moreover, we find a decay timescale $\tau_+ = 43.3 \pm 2.1$ s,  close to the shortest timescales measured in the 2000 and 2004 Vela glitches \citep{dodson+2002, dodson+2007} and within the errors of the value obtained in \citet{ashton+2019}. 
The rise time  $\tau_- = 0.20 \pm 0.14$ s is over two order of magnitude shorter than $\tau_+$; it has quite large errors, reflecting the difficulty to resolve the short time behaviour, but it is well within the upper limit of 12.6 s determined by \citet{ashton+2019}.

The mutual friction parameters $B$ can be directly compared to the constraints given by \citep{graber+2018}, namely $3\times 10^{-5} < \mathcal{B}_{\rm core} < 10^{-4} $ for the  drag between the core superfluid and the normal component, and  $\mathcal{B}_{\rm cr} > 10^{-3} $ for that between the crustal superfluid and the normal component. These values  possibly correspond  to electron scattering off magnetised vortices in the core and kelvon scattering in the crust, the latter parameter being poorly predicted by theory, with differences of more than one order of magnitude at  higher densities between different calculations  \citep{graber+2018}. 

 If we interpret the two superfluid components of our model as the core ($i=1$) and the crustal reservoir ($i=2$), then the value $\mathcal{B}_1=(6.6 \pm 0.6) \cdot 10^{-5}$ lies right in the constrained interval  for $\mathcal{B}_{\rm core}$; the parameter $\mathcal{B}_2$ is affected by a large error (reflecting the large uncertainty of all short time parameters, as seen in Table~\ref{tab:results}) but it also satisfies the lower limit on $\mathcal{B}_{\rm cr}$. 
Since to date calculations of the drag coefficients $\mathcal{R}_i$ are uncertain, the present model provides a simple technique to extract  average values of these parameters from glitch observations, which may help clarifying the theoretical issues concerning the microphysics involved in the dissipative channels at work during a glitch. 

Regarding the time when the glitch begins, $t_{gl}$, our value is before what estimated in \citet{palfreyman+2018}, but within their error bars. We find $t_{gl} \approx t_1$, which supports the idea that the magnetosphere decoupling is associated to the onset of the glitch.

We finally discuss the fractional moments of inertia. In Figure~\ref{fig:MoI} we display the partial fraction of neutrons  for shells starting from the surface and going deeper into the star,  using a unified nucleonic equation of state (SLy4 \citep{sly4}) and for different values of the stellar mass. We  see that the value $x_2 \approx 8\%$ implies that the reservoir cannot be limited to the crust (which contains at most 4\%  of the neutron fraction for the lightest neutron star), but extends into the outer core to densities below nuclear saturation. For a standard 1.4$M_\odot$ star,  the intersection of the curve with the solid horizontal line representing $x_2$ in Figure~\ref{fig:MoI} shows that the reservoir extends to about $0.75 n_0$ (with $n_0 = 0.168 \,$fm$^{-3}$  the nuclear saturation density); this is compatible with some  calculations of S-wave pairing gaps \citep{ho+2015, montoli+2020}. 

We also see that $x_1+x_2 \approx 68\%$, implies that the moment of inertia fraction associated to normal matter is $x_p \approx 32\%$. This is much more than the value predicted by  equations of state without an inner core (between 5\% and 10\%, as shown for SLy4 by the endpoints of the curves  in Figure~\ref{fig:MoI}, which give the total neutron fraction of the star, $x_n$, the remaining $1-x_n$ then being the proton fraction). Therefore, our results  suggest the presence of an inner core of  matter strongly coupled to the charged component. For each mass in Figure~\ref{fig:MoI}, the intersection of the curve with the solid horizontal line  corresponding to $x_1+x_2$ identifies the transition density to the innermost region that is rigidly coupled to the normal component.  For a standard 1.4$M_\odot$ star, such a core would start around $2 n_0$. This is compatible with microscopic calculations, which predict the appearance of an inner core of  non-nucleonic matter (hyperons, meson condensates, quarks) at densities in the range $2n_0 - 3 n_0$.
Other possibilities, however, can be proposed, such as strong coupling of the neutron superfluid  to the proton superconductor in the inner core, due to the (still poorly known) vortex-fluxoid interaction.

\section{Accounting for entrainment}
\label{sec:entrainment}

In this section we introduce  entrainment, namely the non-dissipative coupling between the superfluid and the normal component \citep[see e.g. ][]{haskellsedrakian2017,chamel_hydro_rev}. 
This can be represented by a dimensionless effective mass $m^*$ of the free neutrons.  
The superfluid angular momentum for rigid rotation is given by a mixing between the superfluid and normal component $J_n = I_n ( m^* \Omega_n + (1-m^*)  \Omega_p)$, see e.g. \cite{andersson_slow2001} and  \cite{chamelcarter2006}, and the dynamical equations~\eqref{eq:3c} become:
\begin{align}
\begin{split}
& \dot{\Omega}_p = - \frac{1}{x_p} \left( x_1 \dot{\Omega}_1 + x_2 \dot{\Omega}_2 + |\dot{\Omega}_\infty| \right)
\\
& m^*_1 \dot{\Omega}_1 +(1-m^*_1) \dot{\Omega}_{p}  = -x_p \frac{\Omega_{1p}}{\tau_{1}} 
\\
& m^*_2 \dot{\Omega}_2 +(1-m^*_2) \dot{\Omega}_{p}  = -x_p \frac{\Omega_{2p}}{\tau_{2}} 
\end{split}
\label{eq:3c entr}
\end{align}
where $m^*_{1,2}$ are the (averaged) effective masses for entrainment for the two superfluid components.  The RHS of the equations for the superfluid in \eqref{eq:3c entr} are not effected by entrainment: this approximation holds under the same conditions valid for equations \eqref{eq:drag equation} and \eqref{eq:timescales}, namely that the lags between the superfluids and the normal component are much smaller than the  angular velocity of the normal component  (cf. equation (52) for the vorticity density in \citet{sidery_glitch_2010}, which reduces to $ 2 \Omega_i \simeq 2 \Omega_p(0) \ (i=1,2)$ when $\Omega_{ip} << \Omega_p(0)$). Under such conditions,  the (approximate) relation \eqref{eq:timescales} still holds also in the presence of entrainment. 

To solve the system \eqref{eq:3c entr} we introduce an auxiliary angular velocity, $\Omega_v$, directly related to the vortex density by the Feynman-Onsager relation, and we properly rescale the moments of inertia and the mutual friction coefficient with the effective mass. 
In this way the rescaled dynamical equations in the v-formalism are identical to those in the n-formalism without entrainment \citep{antonellipizzochero2017}. 
In the case of our model with three rigid components the $\Omega_{vi}$ are given by
\begin{align}
\Omega_{vi} = m^*_i \Omega_i +(1-m^*_i)\Omega_p \qquad (i=1,2) \, ,
\label{eq:omegav}
\end{align}
which implies:
\begin{align}
\tilde{\Omega}_{ip} =   \Omega_{vi}-\Omega_{p}  = m^*_i \Omega_{ip}  \qquad (i=1,2).
\label{eq:omegatilde}
\end{align}
The  rescaled (tilded) variables are defined by:
\begin{align}
\tilde{x}_{i} &=  \frac{x_i}{m^*_i} \qquad (i=1,2) 
\\
\tilde{x}_p &=1-\tilde{x}_1-\tilde{x}_2 = x_p-(1-m^*_1)\tilde{x}_1-(1-m^*_2)\tilde{x}_2  
\\
\tilde{\mathcal{B}}_{i} &=  \frac{\mathcal{B}_i}{m^*_i} \qquad (i=1,2) 
\\
\tilde{\tau}_{i} &=  \frac{\tau_i m^*_i \tilde{x}_p}{x_p} = \frac{\tilde{x}_p}{2 \Omega_p(0) \tilde{\mathcal{B}}_i} \qquad (i=1,2) 
\label{eq:renorm}
\end{align}
where we used equation~\eqref{eq:timescales}. By direct substitutions of equations~\eqref{eq:omegav}-\eqref{eq:renorm} in the system of equations~\eqref{eq:3c entr} and after some calculations we finally obtain:
\begin{align}
\begin{split}
\dot{\Omega}_p &= - \frac{1}{\tilde{x}_p} \left( \tilde{x}_1 \dot{\Omega}_{v1} + \tilde{x}_2 \dot{\Omega}_{v2} + |\dot{\Omega}_\infty| \right)
\\
\dot{\Omega}_{v1} &=   -\tilde{x}_p \frac{\tilde{\Omega}_{1p}}{\tilde{\tau}_{1}} 
\\
\dot{\Omega}_{v2} &=   -\tilde{x}_p \frac{\tilde{\Omega}_{2p}}{\tilde{\tau}_{1}} 
\end{split}
\label{eq:3c vform}
\end{align}
which is identical to the system of equations~\eqref{eq:3c}, but for the tilded variables. 

It follows that, in the presence of entrainment, the timing solutions are still represented by equations ~\eqref{eq:solution} and \eqref{eq:res} for the glitch and its residuals, but with tilded parameters instead of  untilded ones. Therefore we do not need to repeat the fit: all the results reported in Table~\ref{tab:results} are still valid, but they now represent  the rescaled quantities. We can then go back to the physical variables using the previous relations: of course, the 'observable' parameters (rise and decay timescale of the overshoot, amplitudes of the exponentials, $\Delta \Omega_p^\infty$, $t_{gl}$ and $r_0$) remain the same, while only the 'internal' parameters (fractional moment of inertia and  mutual friction coefficients) must be rescaled.

For example, we consider the case of no entrainment in the core component and strong entrainment in the reservoir; this is justified by some theoretical calculation, which suggest an effective mass slightly smaller than 1 in the core \citep{chamelhaensel2006} and quite large in the crust \citep{chamel2012}. 
In particular, we take $m^*_1=1$ and $m^*_2=4$, the latter being close to the average value of 4.3-4.3 \citep{andersson+2012,chamel2013}, but other values could be tested: to date, the issue of strong entrainment in the crust is still open to debate \citep{chamel2012, martinurban2016, watanabepethick2017, sauls+2020}. 

In Table~\ref{tab:results_entr} we report the physical quantities whose values are changed because of entrainment, namely the fractional moments of inertia and the mutual friction coefficients; with entrainment being confined to the crust ($i=2$), only the values of the reservoir are affected, namely $\mathcal{B}_2=m^*_2 \tilde{\mathcal{B}}_2$ and $x_2=m^*_2 \tilde{x}_2$. In particular, the value of $\mathcal{B}_2 =(1.1 \pm 0.9) \cdot 10^{-1}$ is four times larger than before and still satisfies the constraint of \citet{graber+2018}; due to the mentioned uncertainty of theoretical calculations, no strong conclusion can be drawn at this stage. As for the fractional moments of inertia, the normal component now results  $x_p \approx 8\%$, in agreement with standard neutron star models without an exotic inner core (indeed, in Figure~\ref{fig:MoI} the dashed horizontal line corresponding to $x_1+x_2=1-x_p$ is very close to the endpoints of the curves for the neutron fraction).  On the other hand, now the reservoir is $x_2 \approx 32\%$, a very  large fraction extending into the outer core up to densities above nuclear saturation. For a standard 1.4$M_\odot$ star,  the intersection of the curve with the dashed horizontal line representing $x_2$ in Figure~\ref{fig:MoI} shows that the reservoir extends to about $1.25 n_0$. This suggests strong non-crustal  pinning,  possibly with the pasta phase and/or the magnetic fluxoids in the superconducting core, but other mechanisms could be envisaged.

\begin{table}
\caption{Fractional moments of inertia and drag parameters obtained in the presence of strong entrainment in the reservoir ($m^*_1=1$ and $m^*_2=4$). The quantities and their propagated errors were obtained by rescaling the results of Table~\ref{tab:results}, as explained  in the text.}
\label{tab:results_entr}
\centering
\begin{tabular}{l @{\hspace{0.1\textwidth}} r}
\hline \hline
Parameter &  Value\\
\hline
$x_1$ & $0.60 \pm 0.01$\\
$x_2$ & $0.32 \pm 0.04$\\
$x_p$ & $0.08 \pm 0.05$  \\
$\mathcal{B}_1$ & $(6.6 \pm 0.6) \cdot 10^{-5}$ \\
$\mathcal{B}_2$ & $(1.1 \pm 0.9) \cdot 10^{-1}$ \\
\hline
\end{tabular}
\end{table}

\section{Conclusions}
\label{sec:conclusion}

We have presented the explicit, analytical timing solution for the minimal three-component model, which confirms the presence of an overshoot when the coupling timescales of the angular momentum reservoir are shorter than those of the superfluid core.

The fit of the 2016 Vela glitch with this  model has provided several interesting physical quantities, like the rise and decay timescales of the overshoot, the time and amplitude of the glitch, and the fractional moments of inertia of the different components. We have compared our results with existing constraints derived from the 2016 Vela glitch, and found agreement with them. 

We have studied the cases with and without strong entrainment in the crustal reservoir: in the former scenario, we find evidence of an inner core strongly coupled to the observable normal component and a reservoir extending beyond  the crust up to densities below nuclear saturation; in the second scenario, the normal component has  standard values of fractional moment of inertia, but the reservoir extends deeper into the outer core, up to densities above nuclear saturation. 

The explicit mathematical form of our model allows to extract physical parameters of the neutron star  directly from well resolved (pulse to pulse) glitch observations in a reasonably simple way.  This may help clarifying some presently open issues, like entrainment in the crust, mutual friction parameters, pinning in the pasta phase and vortex-fluxoids interaction \citep{sourieMNRASLett2020}. 

It would also be interesting to study  the possibility of both components being ``active'' (two distinct reservoirs of angular momentum), as well as to incorporate general relativistic corrections to the moments of inertia \citep{andersson_slow2001,antonelli+2018} and to the timescales \citep{sourieRise2017,gavassino_timescale2020}: we plan to address these issues in future work.

\begin{acknowledgements}
Partial support comes from PHAROS, COST Action CA16214 and INFN.
Marco Antonelli acknowledges support from the Polish National Science Centre grant SONATA BIS 2015/18/E/ST9/00577 (PI: B. Haskell). We thank the anonymous referee  and Brynmor Haskell for important suggestions which have improved our
work.
\end{acknowledgements}

\bibliographystyle{aa}
\bibliography{biblio}

\end{document}